\documentstyle[prd,preprint,tighten,aps,eqsecnum,amssymb,epsfig]{revtex}
\begin{document}
\draft
\preprint{
\begin{tabular}{r}
DFTT 30/95
\\
JHU-TIPAC 95017
\\
hep-ph/9505301
\end{tabular}
}
\title{ATMOSPHERIC NEUTRINO OSCILLATIONS
AMONG THREE NEUTRINO FLAVORS
AND LONG-BASELINE EXPERIMENTS}
\author{
S.M. Bilenky$^{\mathrm{a,b}}$,
C. Giunti$^{\mathrm{a}}$
and
C. W. Kim$^{\mathrm{c}}$
}
\address{
\begin{tabular}{c}
$^{\mathrm{a}}$INFN, Sezione di Torino
and Dipartimento di Fisica Teorica, Universit\`a di Torino,
\\
Via P. Giuria 1, 10125 Torino, Italy
\\
$^{\mathrm{b}}$Joint Institute of Nuclear Research, Dubna, Russia
\\
$^{\mathrm{c}}$ Department of Physics and Astronomy,
The Johns Hopkins University,
\\
Baltimore, Maryland 21218, USA.
\end{tabular}
}
\date{\today}
\maketitle
\begin{abstract}
Anticipating  future long-baseline
neutrino experiments
to search for neutrino oscillations
in various oscillation channels,
we have analyzed the atmospheric neutrino data
in the framework of a model
with mixing of three massive neutrino fields
that can accommodate the data of solar neutrino experiments.
The oscillations of atmospheric and terrestrial neutrinos
are described in this model by three parameters:
$ \Delta m^2 $
and
the squared moduli of two elements of the mixing matrix,
$ | U_{e3} |^2 $
and
$ | U_{\mu3} |^2 $,
which determine the
amplitudes of  oscillations among all
active neutrinos.
The results of the analysis of the atmospheric neutrino
data are presented in the form of allowed regions in
the plane of the two parameters
oscillation amplitude and $ \Delta m^2 $
for all oscillation channels to be investigated in the future long-baseline
oscillation experiments.
\end{abstract}

\pacs{}

\narrowtext

\section{Introduction}
\label{INTRO}

The question of neutrino mass and mixing
is  the central issue in  modern neutrino physics.
The search for the effects of neutrino mass and mixing and
the investigation of the nature of massive neutrinos
(Dirac or Majorana?) may provide an important clue to
explore and understand the physics beyond the standard model.

Important indications in favor of neutrino mixing
have emerged from the solar neutrino data.
In all four solar neutrino experiments
\cite{HOMESTAKE,KAMIOKANDE,GALLEX,SAGE},
that are sensitive to different parts of
the solar neutrino spectrum,
the observed event rates
are significantly less than the event rates predicted by the
standard solar model.
All the existing solar neutrino data can be described with the assumption
that neutrinos are mixed and the resonant MSW mechanism takes
place in the Sun
\cite{MSW}.
For the two parameters
$ \Delta m^2 $ and $ \sin^2 2 \vartheta $
($ \Delta m^2 \equiv m_2^2 - m_1^2 $,
$m_1$ and $m_2$ being the neutrino masses,
and $\vartheta$  the mixing angle)
the following values were obtained:
\begin{eqnarray}
\Delta m^2 \simeq 5 \times 10^{-6} \, \mbox{eV}^2
\quad
\mbox{and}
&
\quad
\sin^2 2\vartheta \simeq 8 \times 10^{-3}
\quad
&
\mbox{(small mixing angle solution)}
\;,
\label{E:SMAMSW}
\\
\Delta m^2 \simeq 10^{-5} \, \mbox{eV}^2
\quad
\mbox{and}
&
\quad
\sin^2 2\vartheta \simeq 0.8
\quad
&
\mbox{(large mixing angle solution)}
\;.
\label{E:LMAMSW}
\end{eqnarray}

Other indications in favor of neutrino mixing
come from the so called atmospheric neutrino anomaly.
The flux ratio of the atmospheric $\nu_\mu$ and $\nu_e$ is close to 2
and is predicted with accuracy better than 5\%
\cite{GAISSERSTANEV,NAUMOV,HONDA,PERKINS}.
For the ratio of ratios
\begin{equation}
R
\equiv
{\displaystyle
N_\mu^{\mathrm{exp}} / N_e^{\mathrm{exp}}
\over\displaystyle
N_\mu^{\mathrm{MC}} / N_e^{\mathrm{MC}}
}
\;,
\label{E050}
\end{equation}
the Kamiokande collaboration
\cite{KAM88,KAM92,KAM94}
found the following value:
\begin{equation}
R
=
0.60^{+0.06}_{-0.05} \pm 0.05
\;.
\label{E:KAMIOKANDE}
\end{equation}
Here,
$ N_{e}^{\mathrm{exp}} $
and
$ N_{\mu}^{\mathrm{exp}} $
are,
respectively,
the number
of observed $e$-like and $\mu$-like events
and
$ N_{e}^{\mathrm{MC}} $
and
$ N_{\mu}^{\mathrm{MC}} $
are,
respectively,
the number
of $e$-like and $\mu$-like events
predicted by MC simulations,
after passing through the same analysis chain as the data
under the assumption that there are no neutrino oscillations.

The atmospheric neutrino anomaly
has also been found  by the IMB
\cite{IMB}
and
Soudan 2
\cite{SOUDAN}
experiments.
In these experiments the following values of
the ratio $R$ have been obtained:
\begin{eqnarray}
R
=
0.54 \pm 0.05 \pm 0.12
&
\null \hskip1cm \null
&
\mbox{(IMB)}
\;,
\label{E:IMB}
\\
R
=
0.64 \pm 0.17 \pm 0.09
&
\null \hskip1cm \null
&
\mbox{(Soudan 2)}
\;.
\label{E:SOUDAN}
\end{eqnarray}

On the other hand,
no indications in favor of
the anomaly
were obtained in the Frejus
\cite{FREJUS89,FREJUS90,FREJUS95}
and
NUSEX
\cite{NUSEX}
experiments. In these experiments the following
values for the ratio $R$ have been obtained:
\begin{eqnarray}
R
=
0.99 \pm 0.13 \pm 0.08
&
\null \hskip1cm \null
&
\mbox{(Frejus)}
\;,
\label{E:FREJUS}
\\
R
=
1.04 \pm 0.25 \phantom{ \null \pm 0.08 }
&
\null \hskip1cm \null
&
\mbox{(NUSEX)}
\;.
\label{E:NUSEX}
\end{eqnarray}
We also  note  that
no indications
in favor of a deficit of muon neutrinos
were found
in the experiments on the detection of
upgoing muons.
For the ratio $r$ of the number of
observed $\mu$-like events
to the expected number of events
(which is much more model dependent
than the ratio of ratios $R$)
the following values were obtained
\cite{KAMUP,IMBUP,BAKSAN,MACRO}:
\begin{eqnarray}
r
=
1.03 \pm 0.04
&
\null \hskip1cm \null
&
\mbox{(IMB)}
\;,
\label{E:IMBUP}
\\
r
=
0.94 \pm 0.06
&
\null \hskip1cm \null
&
\mbox{(Kamiokande)}
\;,
\label{E:KAMUP}
\\
r
=
1.13 \phantom{ \null \pm 0.06 }
&
\null \hskip1cm \null
&
\mbox{(Baksan)}
\;,
\label{E:BAKSAN}
\\
r
=
0.73 \pm 0.16
&
\null \hskip1cm \null
&
\mbox{(MACRO)}
\;.
\label{E:MACRO}
\end{eqnarray}

The Kamiokande,
IMB and Soudan 2 data can be explained by neutrino
oscillations.
{}From the analysis of the Kamiokande data in the simplest case
of oscillations between two types of neutrinos
the following allowed ranges of
the parameters
$ \Delta m^2 $
and
$ \sin^2 2\vartheta $
($\vartheta$ is the mixing angle)
have been found
\cite{KAM94}:
\begin{equation}
5 \times 10^{-3}
\le
\Delta m^2
\le
3 \times 10^{-2} \, \mbox{eV}^2
\quad \quad
0.7
\le
\sin^2 2\vartheta
\le
1
\label{E072}
\end{equation}
in the case of
$ \nu_{\mu} \leftrightarrows \nu_{\tau} $
oscillations
and
\begin{equation}
7 \times 10^{-3}
\le
\Delta m^2
\le
8 \times 10^{-2} \, \mbox{eV}^2
\quad \quad
0.6
\le
\sin^2 2\vartheta
\le
1
\label{E071}
\end{equation}
in the case of
$ \nu_{\mu} \leftrightarrows \nu_{e} $
oscillations.

It is very important to note that
the indications in favor of neutrino mixing
coming from the atmospheric neutrino data
can be checked in future experiments with terrestrial neutrinos.
At present a wide range of long-baseline
experiments with neutrinos from reactors and
accelerators aimed to investigate neutrino oscillations
in the range
$
\Delta m^2
\simeq
10^{-3} - 10^{-2} \, \mbox{eV}^2
$
is under development
(see Ref.\cite{LBNO}).
In these experiments, different neutrino transitions
will be investigated such as
$ \bar\nu_{e} \to \bar\nu_{e} $
with reactors
\cite{CHOOZ}
and
$ \nu_{\mu} \to \nu_{\mu} $,
$ \nu_{\mu} \to \nu_{e} $,
$ \nu_{\mu} \to \nu_{\tau} $
with accelerators
\cite{KEK,E889,MINOS,ICARUS}.
In the accelerator experiments, neutral-current processes will also be
investigated.

In order to accommodate the solar neutrino data and
the atmospheric neutrino anomaly,
corresponding to different scales for the
neutrino mass squared difference,
one must assume
that the flavor neutrino fields are combinations of
at least three fields of massive neutrinos.
We will analyze here the atmospheric neutrino data
in the framework of a minimal scheme with three massive neutrinos
(see also Ref.\cite{LISI}).
In this scheme
the oscillations of atmospheric and terrestrial neutrinos
are characterized by three parameters,
$ \Delta m^2 \equiv m_3^2 - m_1^2 $
($\nu_3$ is the heaviest neutrino)
and the squared moduli of two elements of the mixing matrix,
$ | U_{e3} |^2 $
and
$ | U_{\mu3} |^2 $.

We have found
that the atmospheric neutrinos data are well described by this model.
In presenting the results of our analysis,
we have taken into account the constraints
that follow from reactor and accelerator experiments
which have searched for neutrino oscillations.
We have obtained
the allowed regions of the oscillation
parameters for different oscillation channels
that will be investigated in future neutrino long-baseline
experiments.
The results of our analysis are presented in the form
convenient for a direct comparison with the data of these
experiments.

In Section \ref{MODEL} we  describe briefly the
model under consideration.
In Section \ref{METHOD} and Section \ref{RESULTS} we  present the
method and the results of
the analysis, respectively.

\section{A model with mixing of three massive neutrino fields}
\label{MODEL}

We  assume that the fields of flavor neutrinos
$\nu_{\alpha L}$
(with $\alpha=e,\mu,\tau$)
are superpositions of the left-handed components of
massive neutrino fields given by
\begin{equation}
\nu_{\alpha L}
=
\sum_{k}
U_{\alpha k}
\nu_{kL}
\qquad
(\alpha=e,\mu,\tau)
\;,
\label{E070}
\end{equation}
where $\nu_k$ is the field of a neutrino (Dirac or Majorana)
with mass $m_k$ and $U$ is a unitary mixing matrix.

The number of massive fields in Eq.(\ref{E070})
can range from three to six.
It is to be emphasized that in most models this number is equal to three.
Such mixing can take place
if
the neutrino masses are generated by
the see-saw mechanism
\cite{SEESAW},
or
the neutrino mass term
contains only the left-handed components
of the neutrino flavor fields,
or
the total lepton number is conserved
(see, for example, Refs.\cite{BILENKY,CWKIM}).

In the following we  assume that
the number of massive fields in Eq.(\ref{E070})
is {\em three}.
We enumerate the three neutrino masses in such a way that
$m_1<m_2<m_3$.
The modulus of the amplitude of
$ \nu_{\alpha} \to \nu_{\beta} $
transitions in vacuum can be written in the form
\begin{equation}
\left|
\cal{A}_{\nu_{\alpha}\to\nu_{\beta}}
\right|
=
\left|
\sum_{k=2}^{3}
U_{\beta k}
\left[
\exp
\left(
- i
{\displaystyle
\Delta m^2_{k1} L
\over\displaystyle
2 p
}
\right)
- 1
\right]
U_{\alpha k}^{*}
+
\delta_{\beta\alpha}
\right|
\;.
\label{E031}
\end{equation}
Here
$L$ is the distance between
the neutrino source and detector,
$p$ is the neutrino momentum
and
$ \Delta m^2_{k1} \equiv m^2_k - m^2_1 $.

In order to describe the solar neutrino data and
the atmospheric neutrino anomaly,
it is necessary to assume that
\begin{equation}
\Delta m^2_{21}
\ll
\Delta m^2_{31}
\;.
\label{E101}
\end{equation}
This inequality is in agreement with
the assumption of
a natural hierarchy of neutrino masses,
which is analogous to the mass hierarchy of quarks.

{}From Eqs.(\ref{E031}) and (\ref{E101})
we obtain the following expression
for the probability of
$ \nu_{\alpha} \to \nu_{\beta} $
($ \bar\nu_{\alpha} \to \bar\nu_{\beta} $)
transitions
(with $\beta\not=\alpha$)
of atmospheric and terrestrial neutrinos
\cite{BFP92}:
\begin{equation}
P_{\nu_{\alpha}\to\nu_{\beta}}
=
P_{\bar\nu_{\alpha}\to\bar\nu_{\beta}}
=
{1\over2}
\,
A_{\nu_{\alpha};\nu_{\beta}}
\left(
1
-
\cos
{\displaystyle
\Delta m^2 \, L
\over\displaystyle
2 \, p
}
\right)
\;.
\label{E033}
\end{equation}
Here
$
\Delta m^2
\equiv
\Delta m^2_{31}
$
and
\begin{equation}
A_{\nu_{\alpha};\nu_{\beta}}
=
A_{\nu_{\beta};\nu_{\alpha}}
=
4
\left| U_{\alpha3} \right|^2
\left| U_{\beta3} \right|^2
\label{E073}
\end{equation}
is the
amplitude of
$ \nu_{\alpha} \leftrightarrows \nu_{\beta} $
($ \bar\nu_{\alpha} \leftrightarrows \bar\nu_{\beta} $)
oscillations.

{}From Eq.(\ref{E033})
we obtain the following expression
for the  survival probability
of $\nu_{\alpha}$ ($\bar\nu_{\alpha}$):
\begin{equation}
P_{\nu_{\alpha}\to\nu_{\alpha}}
=
P_{\bar\nu_{\alpha}\to\bar\nu_{\alpha}}
=
1
-
\sum_{\beta\not=\alpha}
P_{\nu_{\alpha}\to\nu_{\beta}}
=
1
-
{1\over2}
\,
B_{\nu_{\alpha};\nu_{\alpha}}
\left(
1
-
\cos
{\displaystyle
\Delta m^2 \, L
\over\displaystyle
2 \, p
}
\right)
\;,
\label{E074}
\end{equation}
where the oscillation amplitude
$ B_{\nu_{\alpha};\nu_{\alpha}} $
is given by
\begin{equation}
B_{\nu_{\alpha};\nu_{\alpha}}
=
\sum_{\beta\not=\alpha}
A_{\nu_{\alpha};\nu_{\beta}}
\;.
\label{E112}
\end{equation}
{}From unitarity of the mixing matrix
we have
\begin{equation}
B_{\nu_{\alpha};\nu_{\alpha}}
=
4
\left| U_{\alpha3} \right|^2
\left(
1
-
\left| U_{\alpha3} \right|^2
\right).
\label{E075}
\end{equation}
In the model under consideration there are 3 free parameters,
$ \Delta m^2 $,
$ \left| U_{e3} \right|^2 $
and
$ \left| U_{\mu3} \right|^2 $.
{}From the unitarity of the mixing matrix we have
$
\left| U_{\tau3} \right|^2
=
1
-
\left| U_{e3} \right|^2
-
\left| U_{\mu3} \right|^2
$.
We will use the expressions in Eqs.(\ref{E033}) and (\ref{E074})
for the analysis of atmospheric neutrino data
and for the discussion of future long-baseline experiments.

In the model under consideration
all oscillation channels are open
and
the probabilities of all possible
transitions of atmospheric and terrestrial
neutrinos have
the same form as in the case of oscillations
between two neutrino types.
This is related to  the fact that all the expressions of the
transition probabilities depend only on one
neutrino mass squared difference
and the oscillation length
$
L_{\mathrm{osc}}
=
4 \pi p / \Delta m^2
$
is the same for all channels.
However,
each oscillation channel is characterized by
its own amplitude.
The oscillation amplitudes
are related by
(see Eq.(\ref{E112}))
\begin{eqnarray}
&&
A_{\nu_{\mu};\nu_{e}}
+
A_{\nu_{\mu};\nu_{\tau}}
=
B_{\nu_{\mu};\nu_{\mu}}
\;,
\label{E113}
\\
&&
A_{\nu_{\mu};\nu_{e}}
+
A_{\nu_{e};\nu_{\tau}}
=
B_{\nu_{e};\nu_{e}}
\;.
\label{E114}
\end{eqnarray}
We add here that
the oscillation amplitudes
$ A_{\nu_{\alpha};\nu_{\beta}} $
satisfy the following relation:
\begin{equation}
\left(
{\displaystyle
A_{\nu_{\mu};\nu_{e}}
A_{\nu_{\mu};\nu_{\tau}}
\over\displaystyle
A_{\nu_{e};\nu_{\tau}}
}
\right)^{1/2}
+
\left(
{\displaystyle
A_{\nu_{\mu};\nu_{e}}
A_{\nu_{e};\nu_{\tau}}
\over\displaystyle
A_{\nu_{\mu};\nu_{\tau}}
}
\right)^{1/2}
+
\left(
{\displaystyle
A_{\nu_{e};\nu_{\tau}}
A_{\nu_{\mu};\nu_{\tau}}
\over\displaystyle
A_{\nu_{\mu};\nu_{e}}
}
\right)^{1/2}
=
2
\;.
\label{E202}
\end{equation}
This relation follows from Eq.(\ref{E073})
and the unitarity of the mixing matrix.

\section{The method of analysis of the data}
\label{METHOD}

In this section we describe our method
of analysis of the experimental data.
As a first step we
have analyzed only the Kamiokande data.
Then, we have carried out
a combined analysis of
the Kamiokande and Frejus data.

The Kamiokande collaboration has measured the ratio of ratios $R$
in two ranges of energies:
the sub-GeV range with energies
of contained events  less than 1.33 GeV and
the multi-GeV range
with energies of contained and partially contained events
more than 1.33 GeV.
The measured values of the ratio $R$
in these two regions agree with each other:
\begin{eqnarray}
&&
R
=
0.60^{+0.06}_{-0.05} \pm 0.05
\hskip0.5cm
\mbox{(sub-GeV energy range)}
\;,
\label{E:SUBGEV}
\\
&&
R
=
0.57^{+0.08}_{-0.07} \pm 0.07
\hskip0.5cm
\mbox{(multi-GeV energy range)}
\;.
\label{E:MULTIGEV}
\end{eqnarray}
The Kamiokande collaboration has also investigated
the dependence of the ratio $R$ on
the zenith-angle $\theta$
($\theta=0$ corresponds
to downward-going neutrinos).
Some indications in favor of a zenith-angle
dependence of the ratio
$R$ were found in the multi-GeV region.

The experimental data
are divided into 5 bins
corresponding to the following averaged values of
$ \cos\theta $:
$ \left\langle \cos\theta \right\rangle_{i=1,\ldots,5} =
-0.8 \, , \, -0.4 \, , \, 0.0 \, , \, 0.4 \, , \, 0.8 $
and the averaged distances
$ \left\langle L \right\rangle_{i=1,\ldots,5} =
10,230 \, , \, 5,157 \, , \, 852 \, , \, 54 \, , \, 26 \, \mbox{Km} $,
respectively.
The number
$ N_{e}^{i} $
of $e$-like events
and
the number
$ N_{\mu}^{i} $
of $\mu$-like events in the bin $i$
are given by
\begin{eqnarray}
&&
N_{e}^{i}
=
N_{e}^{i\mathrm{MC}}
\,
P_{\nu_{e}\to\nu_{e}}^{i}
+
N_{\mu}^{i\mathrm{MC}}
\,
P_{\nu_{\mu}\to\nu_{e}}^{i}
\;,
\label{E011}
\\
&&
N_{\mu}^{i}
=
N_{e}^{i\mathrm{MC}}
\,
P_{\nu_{e}\to\nu_{\mu}}^{i}
+
N_{\mu}^{i\mathrm{MC}}
\,
P_{\nu_{\mu}\to\nu_{\mu}}^{i}
\;.
\label{E012}
\end{eqnarray}
Here
$ N_{e}^{i\mathrm{MC}} $
and
$ N_{\mu}^{i\mathrm{MC}} $
are,
respectively,
the number
of $e$-like and $\mu$-like events
predicted by MC simulations
under the assumption that there are no oscillations
and
$ P_{\nu_{\alpha}\to\nu_{\beta}}^{i} $
is the averaged probability of
$\nu_{\alpha}\to\nu_{\beta}$
transitions
(with $\alpha,\beta=e,\mu$).
{}From Eqs.(\ref{E011}) and (\ref{E012}),
we obtain  the ratio of the ratios
$R_{i}$
in the bin $i$
\begin{equation}
R_{i}
=
{\displaystyle
P_{\nu_{\mu}\to\nu_{\mu}}^{i}
+
\left( N_{\mu}^{i\mathrm{MC}} / N_{e}^{i\mathrm{MC}} \right)^{-1}
P_{\nu_{e}\to\nu_{\mu}}^{i}
\over\displaystyle
P_{\nu_{e}\to\nu_{e}}^{i}
+
\left( N_{\mu}^{i\mathrm{MC}} / N_{e}^{i\mathrm{MC}} \right)
P_{\nu_{\mu}\to\nu_{e}}^{i}
}
\; .
\label{E013}
\end{equation}

In our calculation we
take into account
30\% errors of the MC calculations of the
total number of
$e$-like and $\mu$-like events
due to the fact that different
calculations give different neutrino fluxes.
The errors of the MC calculations of the
$\mu/e$ ratio are much smaller:
9\% for the Kamiokande sub-GeV data,
12\% for the Kamiokande multi-GeV data
and
14\% for the Frejus data.
In our calculations
we have neglected the fact that not all
$e$-like and $\mu$-like events
are produced by
$ \nu_{e} $ ($\bar\nu_{e}$)
and
$\nu_{\mu} $ ($\bar\nu_{\mu}$)
interactions,
respectively.
The purity of the
$e$-like and $\mu$-like events
is estimated by the Kamiokande Collaboration
to be higher than 90\%.

For the fit of the sub-GeV Kamiokande data
we have used
the ratio of ratios
given in Eq.(\ref{E013}).
For the fit of the multi-GeV Kamiokande data
and the Frejus data
we have used
the number of $e$-like and $\mu$-like events
given in Eqs.(\ref{E011}) and (\ref{E012}).
The expected numbers of $e$-like and $\mu$-like events,
$ N_{e}^{i\mathrm{MC}} $
and
$ N_{\mu}^{i\mathrm{MC}} $,
were taken from
Fig.2 of Ref.\cite{KAM88}
for the Kamiokande sub-GeV data,
from Fig.3 of Ref.\cite{KAM94}
for the Kamiokande multi-GeV data
and
from
from Fig.4 of Ref.\cite{FREJUS89}
for the Frejus data, respectively.
The quantities
$ P_{\nu_{e}\to\nu_{\mu}}^{i} $
and
$ P_{\nu_{\mu}\to\nu_{e}}^{i} $
are the
$ \nu_{e}\to\nu_{\mu} $
and
$ \nu_{\mu}\to\nu_{e} $
transition probabilities
averaged over the neutrino energy spectrum
and the zenith-angles of bin $i$.
These probabilities depend on the value of
$ \Delta m^2 $
and
they are different because
the initial $ \nu_{e} $ and $ \nu_{\mu} $
have different energy spectra.
These energy spectra are given
in Fig.1 of Ref.\cite{KAM92}
for the Kamiokande sub-GeV region,
in Fig.2 of Ref.\cite{KAM94}
for the Kamiokande multi-GeV region
and
in Fig.3 of Ref.\cite{FREJUS89}
for the Frejus experiment, respectively.

Since the expressions (\ref{E033}) and (\ref{E074})
that we use for the transition and survival probabilities
of flavor neutrinos
are valid in vacuum,
we do not consider in our analysis the Kamiokande
multi-GeV data on upward-going neutrinos,
for which the matter effect
could be important
(see Ref.\cite{MATTER}).

Finally,
we would like to emphasize
that our analyses
take into account the possibility
of  simultaneous oscillations among
all three neutrino types:
$ \nu_{\mu} \leftrightarrows \nu_{e} $,
$ \nu_{\mu} \leftrightarrows \nu_{\tau} $
and
$ \nu_{e} \leftrightarrows \nu_{\tau} $.
This analysis is quite different
from the usual two generation analyses,
where the two oscillation channels
$ \nu_{\mu} \leftrightarrows \nu_{e} $
and
$ \nu_{\mu} \leftrightarrows \nu_{\tau} $
are considered separately,
leading to two possible unrelated
interpretations of the data.

\section{Results of the analysis}
\label{RESULTS}

The results of the analysis
of the Kamiokande data
in the framework of the
model
with oscillations among
three  neutrino flavors
discussed in Section \ref{MODEL}
are presented in Figs.\ref{FIG1}--\ref{FIG4}.
We have obtained a rather good description of the data, i.e.
for 8 degrees of freedom we obtained
$ \chi^2_{\mathrm{min}} = 3.2 $
at
$ \Delta m^2 = 2.7 \times 10^{-2} \, \mbox{eV}^2 $,
$ \left| U_{e3} \right|^2 = 0.14 $
and
$ \left| U_{\mu3} \right|^2 = 0.62 $
(with a confidence level (CL) of 92\%).
In the figures we present
the results of our analysis
in terms of
two-parameter allowed regions
at 90\% CL
(these regions are given by
$ \chi^2 \le \chi^2_{\mathrm{min}} + 4.6 $).

In Fig.\ref{FIG1} we have shown
the region allowed by the Kamiokande data
in the plane of the parameters
$ B_{\nu_e;\nu_e} $
and
$ \Delta m^2 $.
We have also shown the forbidden regions
(on the right of the corresponding curves)
that were found from the analysis of the data from the
Bugey
\cite{BUGEY}
and
Krasnoyarsk
\cite{KRASNOYARSK}
reactor experiments
and
the region of sensitivity
of the future
CHOOZ
\cite{CHOOZ}
reactor experiment.
It is seen from Fig.\ref{FIG1} that a large part
of the region in the
$ B_{\nu_e;\nu_e} $--$ \Delta m^2 $
plane
that is allowed by the Kamiokande atmospheric neutrino data
will be investigated by future long-baseline reactor
experiments.

In Fig.\ref{FIG2} we have shown
the region allowed by the Kamiokande data
in the
$ B_{\nu_{\mu};\nu_{\mu}} $--$ \Delta m^2 $
plane.
As can be seen from Fig.\ref{FIG2},
only large values of $ B_{\nu_{\mu};\nu_{\mu}} $
($ B_{\nu_{\mu};\nu_{\mu}} \gtrsim 0.5 $)
are allowed by the results of this experiment.
The regions of sensitivity
of the future
BNL E889
\cite{E889},
MINOS
\cite{MINOS}
and
KEK--Super-Kamiokande
\cite{KEK}
long-baseline
experiments are also shown.
It is seen from this figure that a  large part of the region of the
$ B_{\nu_{\mu};\nu_{\mu}} $--$ \Delta m^2 $
plane,
including the Kamiokande-allowed region,
will be investigated in future long-baseline accelerator
disappearance experiments.

The Kamiokande-allowed region in the plane
of the parameters
$ A_{\nu_{\mu};\nu_{e}} $
and
$ \Delta m^2 $
is shown in Fig.\ref{FIG3}.
In this figure
we have also shown the regions that are
forbidden by the results of the Bugey and Krasnoyarsk
reactor experiments
and
the regions of sensitivity of
the future KEK-SK, MINOS
and ICARUS
\cite{ICARUS}
long-baseline experiments
(we have used the inequality
$A_{\nu_{\mu};\nu_{e}}
\le
B_{\nu_{e};\nu_{e}}
$).
Finally,
in Fig.\ref{FIG4}
we present the results of our calculations
of the allowed region in the plane of the parameters
$ A_{\nu_{\mu};\nu_{\tau}} $
and
$ \Delta m^2 $.
The region limited by the solid lines is
allowed by the Kamiokande data alone.
A large part of this
region is excluded by the results of reactor experiments.
In fact,
for relatively small values of the amplitude
$ A_{\nu_{\mu};\nu_{\tau}} $,
the fit
of the Kamiokande data requires large values
of the amplitude
$ A_{\nu_{\mu};\nu_{e}} $,
which,
due to inequality
$ A_{\nu_{\mu};\nu_{e}} \le B_{\nu_{e};\nu_{e}} $,
can be in contradiction
with the results of reactor experiments.
The region in the
$ A_{\nu_{\mu};\nu_{\tau}} $--$ \Delta m^2 $
plane
that is allowed by the Kamiokande data and
by the results of the Bugey and Krasnoyarsk reactor experiments
lies on the right
of the dotted line in Fig.\ref{FIG4}.
As can be seen from Fig.\ref{FIG4},
the Kamiokande atmospheric neutrino data allow
large values of the amplitude of
$ \nu_{\mu} \leftrightarrows \nu_{\tau} $
oscillations and
the values of the
parameter $ \Delta m^2 $ in the range
\begin{equation}
8 \times 10^{-3} \, \mbox{eV}^2
\lesssim
\Delta m^2
\lesssim
8 \times 10^{-2} \, \mbox{eV}^2
\;.
\label{E120}
\end{equation}
For a neutrino energy $ E \simeq 15 \, \mbox{GeV} $,
the range (\ref{E120})
corresponds to the following
range of possible values of the oscillation length:
\begin{equation}
450 \, \mbox{Km}
\lesssim
L_{\mathrm{osc}}
\lesssim
4500 \, \mbox{Km}
\;.
\label{E061}
\end{equation}
Therefore,
a large effect of
$\nu_\mu\to\nu_\tau$
oscillations could be observed
with the long-baseline neutrino oscillation
experiments now under preparation
at CERN and FNAL
with detectors placed
in the Gran Sasso Laboratory
and
in the Soudan Mine,
at a distance of about
730 Km from the corresponding neutrino source.
The best fit of the Kamiokande data corresponds to
$
L_{\mathrm{osc}}
\simeq
1400 \, \mbox{Km}
$.
Thus,
a large fraction of the charged-current events
in long-baseline experiments could be events
in which $\tau$ is produced.

In Fig.\ref{FIG4}
we have also shown the region
in the plane of the parameters
$ A_{\nu_{\mu};\nu_{\tau}} $--$ \Delta m^2 $
that is forbidden by the data of the CDHS experiment
\cite{CDHS}
in which $\nu_\mu$ disappearance was searched for.
It can be seen
that this forbidden region
is far away from the region allowed
by the Kamiokande data.

Up to now we have discussed
the results
of the analysis of the data of the Kamiokande experiment.
It is interesting to see
what happens to the quality of the  fit as well as
to the allowed regions
if we make a combined analysis
of the Kamiokande data,
which indicate  neutrino oscillation effects,
and the Frejus data,
which are not in contradiction
with the absence of neutrino oscillations.

We have found that
the Kamiokande and Frejus data
can simultaneously be explained in the framework
of our three parameters model
with mixing among three neutrinos.
For
18 degrees of freedom
we obtained
$ \chi^2_{\mathrm{min}} = 22 $
at
$ \Delta m^2 = 1.4 \times 10^{-2} \, \mbox{eV}^2 $,
$ \left| U_{e3} \right|^2 = 0.06 $
and
$ \left| U_{\mu3} \right|^2 = 0.23 $
(with a CL of 22\%).
The results of the combined fit
are presented in Figs.\ref{FIG5}--\ref{FIG8}.
In each of these figures
we have shown
the regions in the plane of two parameters
which is allowed both by Kamiokande
and Frejus data
at 90\% CL.
We have also shown
the regions forbidden by the data
of the reactor and accelerator experiments
and the regions of sensitivity
of the future long-baseline experiments.

In conclusion,
we would like to make the following remark:
in order to accommodate the solar neutrino data
it is necessary to require that the parameter
$ \left| U_{e3} \right| $
is not too large.
In fact,
the survival probability of solar neutrinos
is given by
\cite{SOLARTHREEGEN}
\begin{equation}
P_{\nu_e\to\nu_e}(E)
=
\left(
1
-
\left| U_{e3} \right|^2
\right)^2
P_{\nu_e\to\nu_e}^{(1,2)}(E)
+
\left| U_{e3} \right|^4
\;,
\label{E172}
\end{equation}
where
$ P_{\nu_e\to\nu_e}^{(1,2)}(E) $
is the survival probability
due to the mixing between
the first and the second generations.
{}From Eq.(\ref{E172}) it is clear that
$ P_{\nu_e\to\nu_e}(E) \ge \left| U_{e3} \right|^4 $
for all values of $E$.
On the other hand,
from a model independent analysis of the solar neutrino data
(see Ref.\cite{BAHCALL94})
it follows that
for $^{7}\mathrm{Be}$ neutrinos
$ P_{\nu_e\to\nu_e} \lesssim 0.5 $.
Therefore,
large values of
$ \left| U_{e3} \right| $
are excluded by the solar neutrino data.
We have first performed the analysis of the experimental data without
any constraint on the value of
$ \left| U_{e3} \right| $.
Then the analysis was repeated
with the constraints
$ \left| U_{e3} \right|^2 \le 0.7 $
and
$ \left| U_{e3} \right|^2 \le 0.5 $.
However,
the inclusion of these constraints did not
induce any modification
of the allowed
regions of the parameters
(this is due to the fact that
all the allowed values of the oscillation amplitudes
can be obtained with
$ \left| U_{e3} \right|^2 \le 0.5 $).
Therefore,
all the allowed regions
shown in the figures
are consistent with the solar neutrino data.

\section{Conclusions}
\label{CONCLUSIONS}

A wide range of reactor and accelerator
long-baseline neutrino experiments which will
investigate different channels of
neutrino oscillations in the range
$ \Delta m^2 \simeq 10^{-3} - 10^{-2} \, \mbox{eV}^2 $
and a broad range
of oscillation amplitudes is
at present under development
at Fermilab, CERN and other laboratories.
Taking into account these new possibilities
for the investigation of neutrino mixing,
we have presented here the results of an analysis of the data
of the underground experiments on the detection
of the atmospheric neutrinos,
in some of which
(Kamiokande, IMB, Soudan 2)
positive indications
in favor of neutrino oscillations with
were found.

The analysis has been carried out by using
the framework of a model with mixing of three
massive neutrino fields which permit simultaneous
oscillations among all active neutrinos.
We have assumed
that there are two scales of
neutrino mass squared difference
which can accommodate the solar neutrino problem
(at a scale of about $ 10^{-5} \, \mbox{eV}^2 $)
and the atmospheric neutrino
anomaly
(at a scale of about $ 10^{-2} \, \mbox{eV}^2 $).
In this model
all oscillation channels for
atmospheric and terrestrial neutrinos are
characterized by the same
oscillation length
$
L_{\mathrm{osc}}
=
4 \pi p / \Delta m^2
$
and
the oscillation amplitudes in different channels
are determined by the two parameters
$ | U_{e3} |^2 $
and
$ | U_{\mu3} |^2 $.
Due to the unitarity of the mixing matrix,
different oscillation amplitudes are connected
by the relations
(\ref{E113})--(\ref{E202}).

The results of the analysis of
the atmospheric neutrino
data are presented in the form of plots
in the plane of the two parameters
oscillation amplitude and $ \Delta m^2$.
In our analysis we have considered the following channels:
$ \nu_\mu \leftrightarrows \nu_x $,
$ \nu_e \leftrightarrows \nu_x $,
$ \nu_\mu \leftrightarrows \nu_e $
and
$ \nu_\mu \leftrightarrows \nu_\tau $.
In the figures we have also presented the limits that were
obtained in reactor and accelerator experiments
searching for neutrino oscillations as well as the
 regions in  different oscillation channels that will be
investigated in future long-baseline experiments.
The results of the analysis
of the Kamiokande data
are shown in Figs.\ref{FIG1}--\ref{FIG4}
and
those of the combined analysis of
the Kamiokande and Frejus data
are shown in Figs.\ref{FIG5}--\ref{FIG8}.

\acknowledgments

It is a pleasure to thank
A. Bottino
for useful discussions.
C.W.K. would like to thank
the Sezione di Torino of INFN
for its hospitality
during the completion of this work.
This work was supported in part by the National Science Foundation.


\begin{figure}[p]
\protect\caption{
Region
of the parameters
$ B_{\nu_e;\nu_e} $
and
$ \Delta m^2 $
allowed at 90\% CL
by the Kamiokande data
(within the two solid lines).
The regions on the right of the dashed and dotted lines
are forbidden by the results of the
Bugey and Krasnoyarsk experiments,
respectively.
The dash-dotted line indicates the sensitivity
of the future long-baseline CHOOZ experiment.
}
\label{FIG1}
\end{figure}

\begin{figure}[p]
\protect\caption{
Region
of the parameters
$ B_{\nu_\mu;\nu_\mu} $
and
$ \Delta m^2 $
allowed at 90\% CL
by the Kamiokande data
(within the solid line).
The region on the right of the short-dashed line
is forbidden by the results of the
CDHS experiment.
The dash-dotted, dash-dot-dotted and long-dashed
lines indicate the sensitivity
of the future long-baseline
BNL E889, MINOS and KEK--Super-Kamiokande
experiments,
respectively.
}
\label{FIG2}
\end{figure}

\begin{figure}[p]
\protect\caption{
Region
of the parameters
$ A_{\nu_\mu;\nu_e} $
and
$ \Delta m^2 $
allowed at 90\% CL
by the Kamiokande data
(within the two solid lines).
The regions on the right of the dashed and dotted lines
are forbidden by the results of the
Bugey and Krasnoyarsk experiments,
respectively.
The dash-dotted, dash-dot-dotted and long-dashed
lines indicate the sensitivity
of the future long-baseline
ICARUS, MINOS and KEK--Super-Kamiokande
experiments,
respectively.
}
\label{FIG3}
\end{figure}

\begin{figure}[p]
\protect\caption{
Region
of the parameters
$ A_{\nu_\mu;\nu_\tau} $
and
$ \Delta m^2 $
allowed at 90\% CL
by the Kamiokande data
(within the two solid lines).
The region on the left of the dotted line
is excluded by the results of the
Krasnoyarsk and Bugey reactor experiments.
The dash-dotted and dash-dot-dotted
lines indicate the sensitivity
of the future long-baseline
ICARUS and MINOS
experiments,
respectively.
}
\label{FIG4}
\end{figure}

\begin{figure}[p]
\protect\caption{
Region
of the parameters
$ B_{\nu_e;\nu_e} $
and
$ \Delta m^2 $
allowed at 90\% CL
by the Kamiokande and Frejus data
(within the solid line).
The regions on the right of the dashed and dotted lines
are forbidden by the results of the
Bugey and Krasnoyarsk experiments,
respectively.
The dash-dotted line indicates the sensitivity
of the future long-baseline CHOOZ experiment.
}
\label{FIG5}
\end{figure}

\begin{figure}[p]
\protect\caption{
Region
of the parameters
$ B_{\nu_\mu;\nu_\mu} $
and
$ \Delta m^2 $
allowed at 90\% CL
by the Kamiokande and Frejus data
(within the solid line).
The region on the right of the short-dashed line
is forbidden by the results of the
CDHS experiment.
The dash-dotted, dash-dot-dotted and long-dashed
lines indicate the sensitivity
of the future long-baseline
BNL E889, MINOS and KEK--Super-Kamiokande
experiments,
respectively.
}
\label{FIG6}
\end{figure}

\begin{figure}[p]
\protect\caption{
Region
of the parameters
$ A_{\nu_\mu;\nu_e} $
and
$ \Delta m^2 $
allowed at 90\% CL
by the Kamiokande and Frejus data
(within the solid line).
The regions on the right of the dashed and dotted lines
are forbidden by the results of the
Bugey and Krasnoyarsk experiments,
respectively.
The dash-dotted, dash-dot-dotted and long-dashed
lines indicate the sensitivity
of the future long-baseline
ICARUS, MINOS and KEK--Super-Kamiokande
experiments,
respectively.
}
\label{FIG7}
\end{figure}

\begin{figure}[p]
\protect\caption{
Region
of the parameters
$ A_{\nu_\mu;\nu_\tau} $
and
$ \Delta m^2 $
allowed at 90\% CL
by the Kamiokande and Frejus data
(within the solid line).
The region on the left of the dotted line
is excluded by the results of the
Krasnoyarsk and Bugey reactor experiments.
The dash-dotted and dash-dot-dotted
lines indicate the sensitivity
of the future long-baseline
ICARUS and MINOS
experiments,
respectively.
}
\label{FIG8}
\end{figure}

\newpage
\setcounter{figure}{0}

\begin{figure}[p]
\begin{center}
\mbox{\epsfig{file=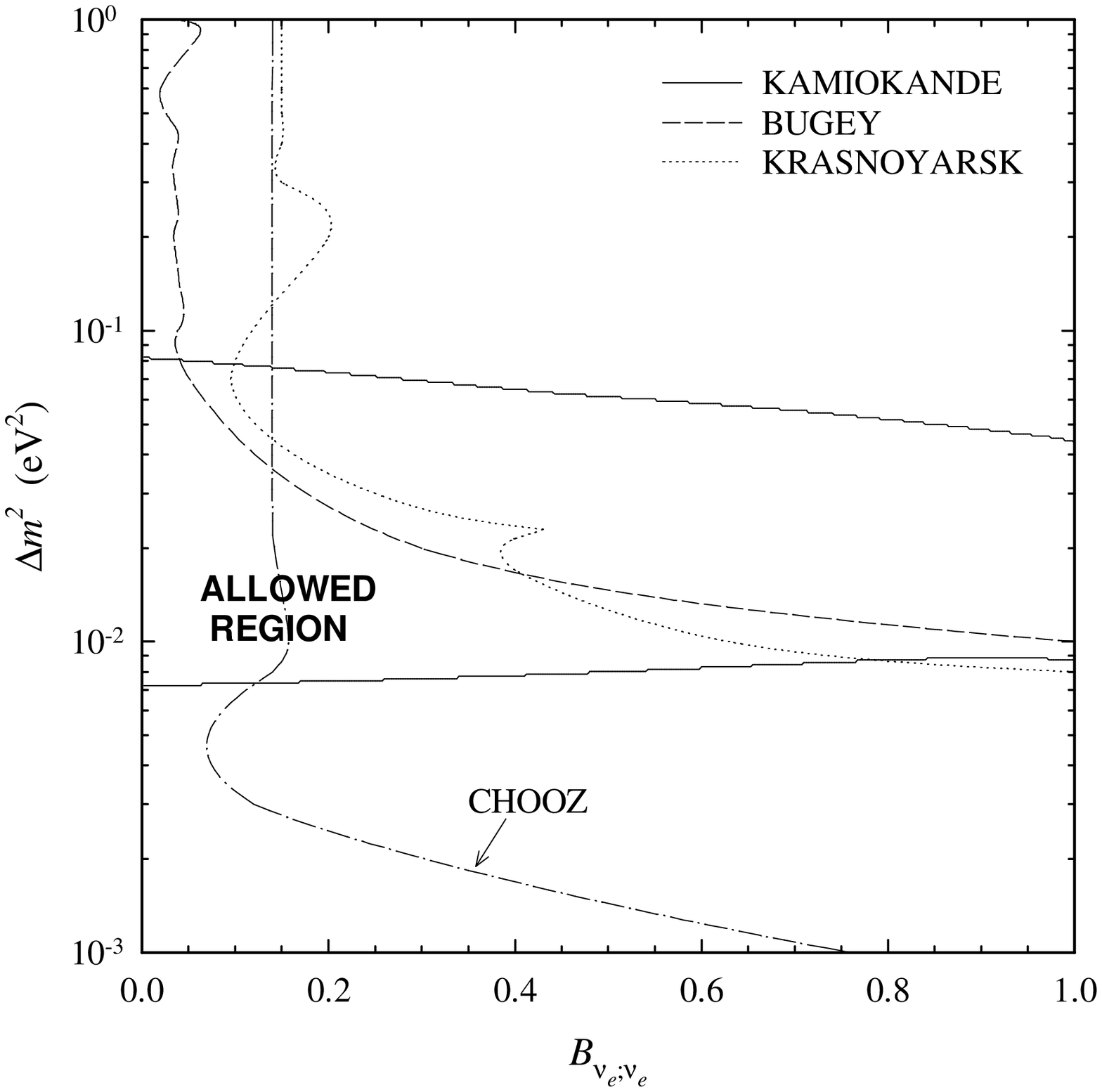,width=\textwidth}}
\end{center}
\vspace{1cm}
\begin{center}
{\Large Figure \ref{FIG1}}
\end{center}
\end{figure}

\begin{figure}[p]
\begin{center}
\mbox{\epsfig{file=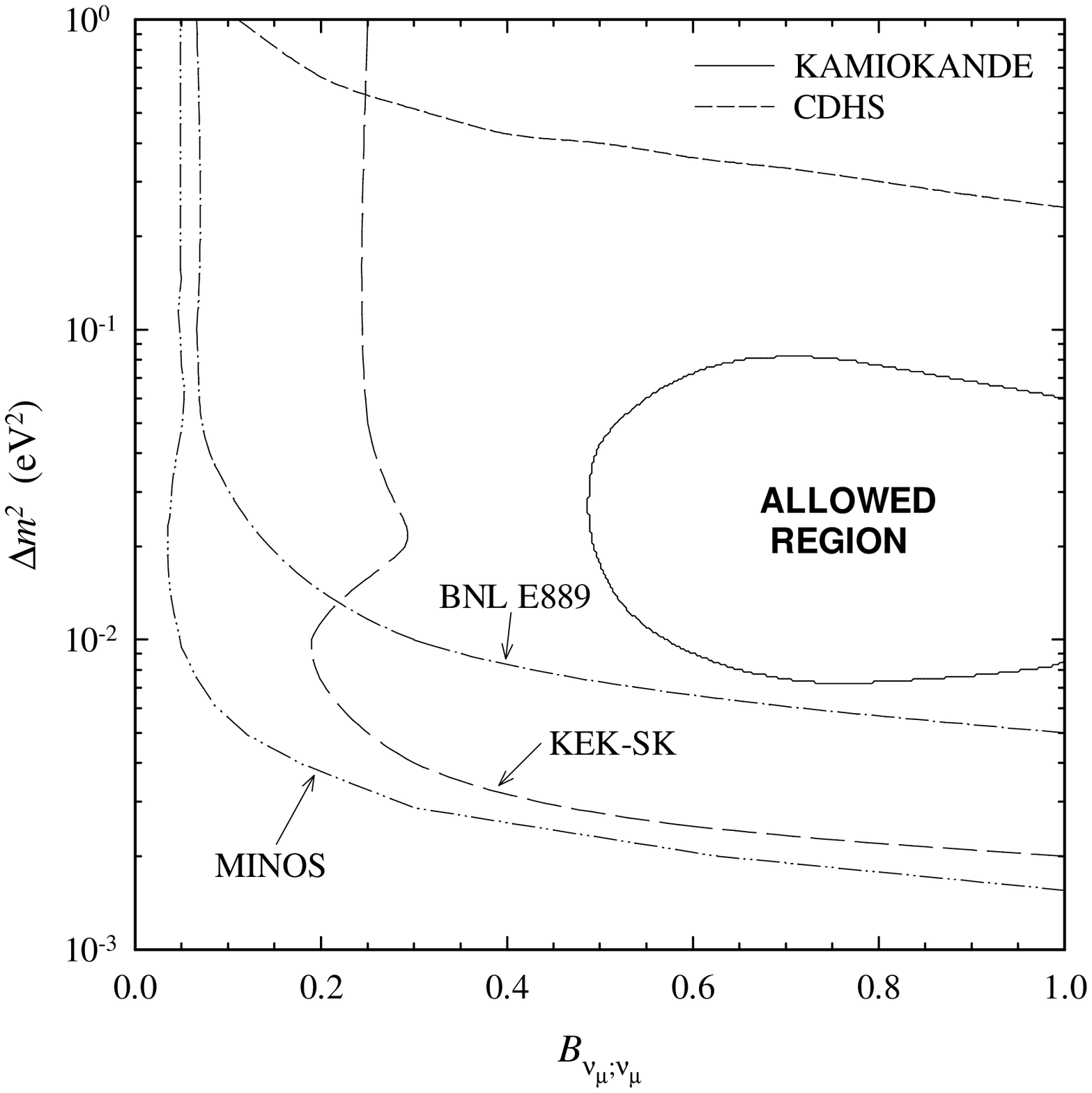,width=\textwidth}}
\end{center}
\vspace{1cm}
\begin{center}
{\Large Figure \ref{FIG2}}
\end{center}
\end{figure}

\begin{figure}[p]
\begin{center}
\mbox{\epsfig{file=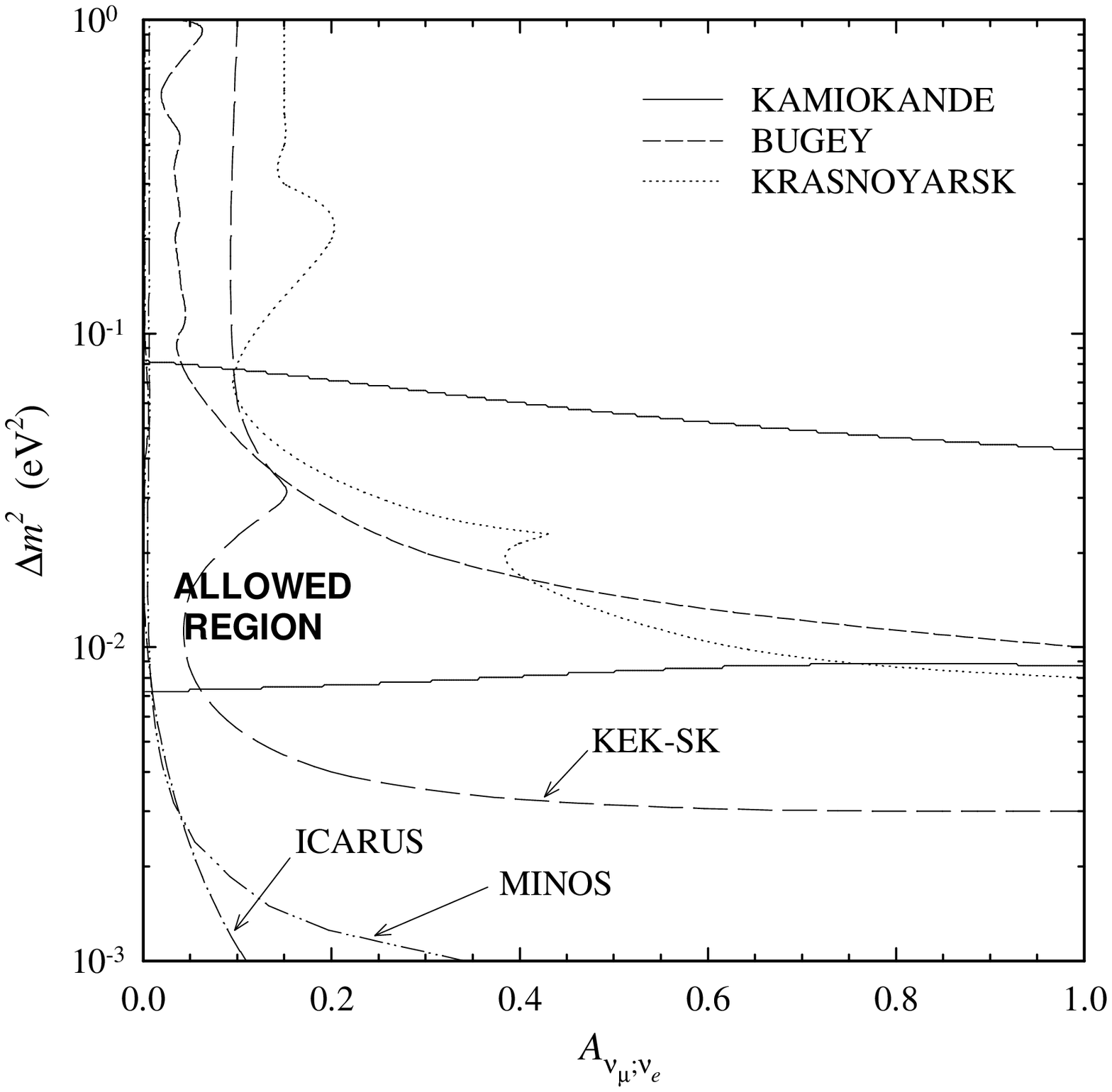,width=\textwidth}}
\end{center}
\vspace{1cm}
\begin{center}
{\Large Figure \ref{FIG3}}
\end{center}
\end{figure}

\begin{figure}[p]
\begin{center}
\mbox{\epsfig{file=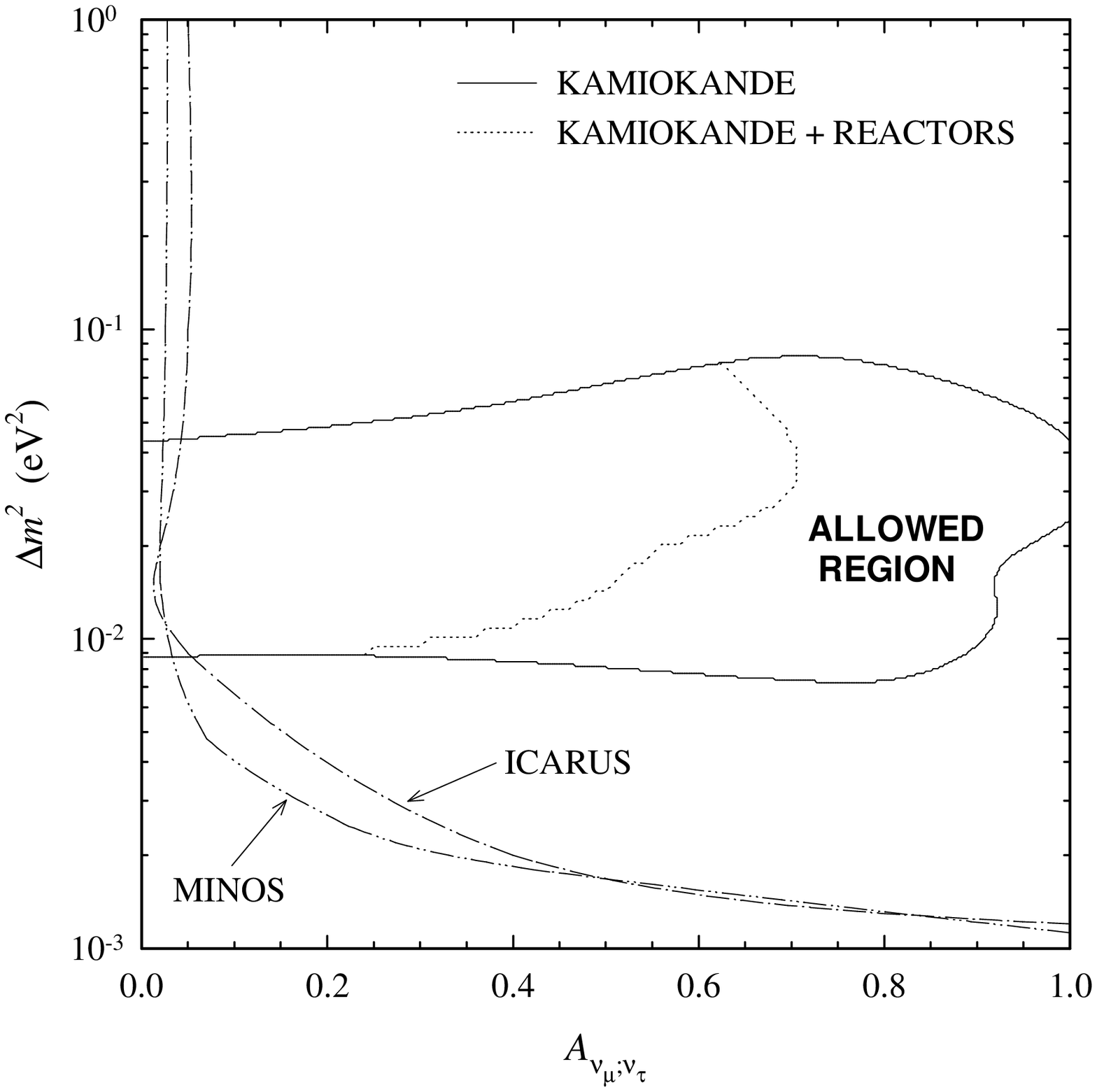,width=\textwidth}}
\end{center}
\vspace{1cm}
\begin{center}
{\Large Figure \ref{FIG4}}
\end{center}
\end{figure}

\begin{figure}[p]
\begin{center}
\mbox{\epsfig{file=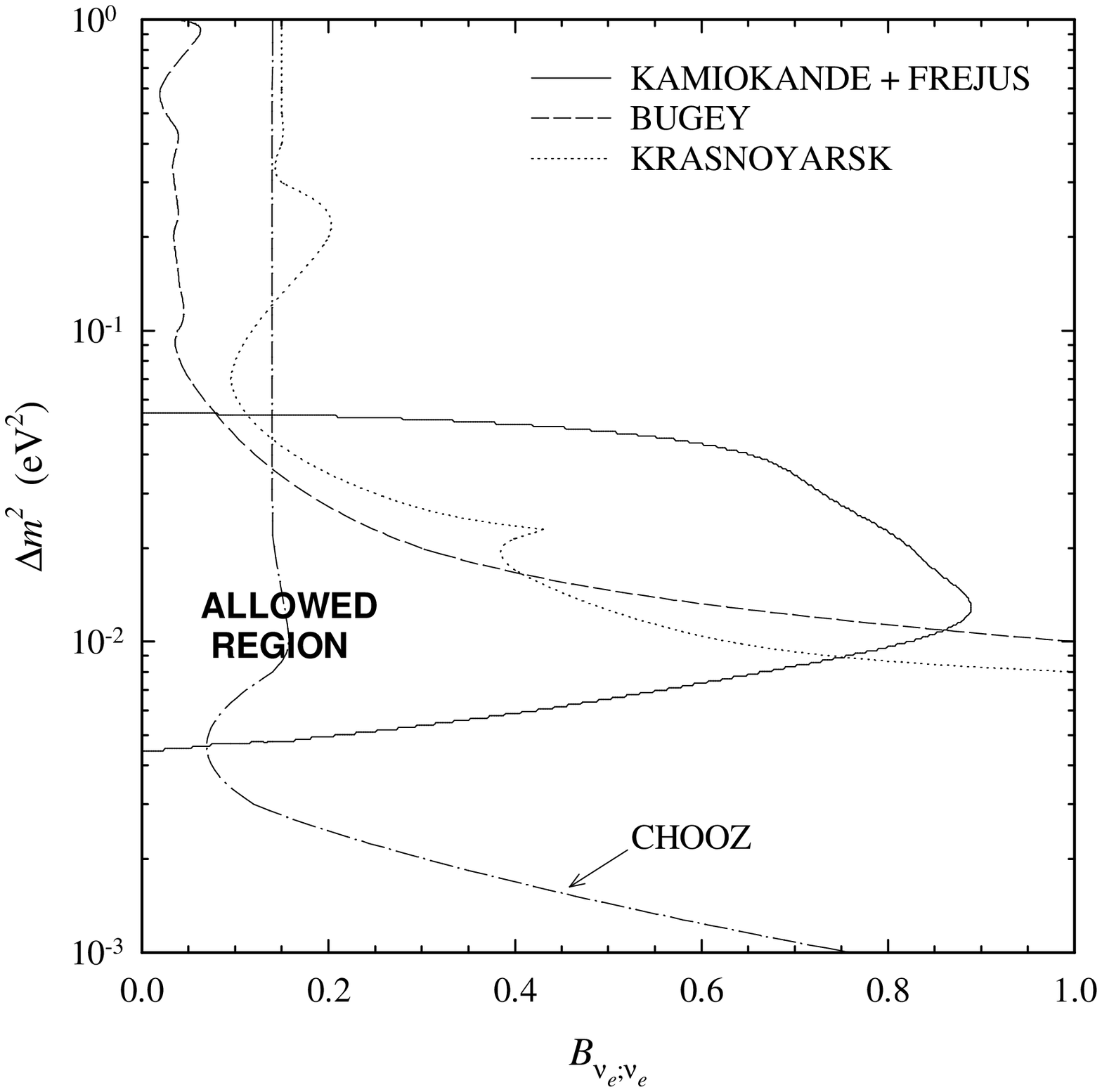,width=\textwidth}}
\end{center}
\vspace{1cm}
\begin{center}
{\Large Figure \ref{FIG5}}
\end{center}
\end{figure}

\begin{figure}[p]
\begin{center}
\mbox{\epsfig{file=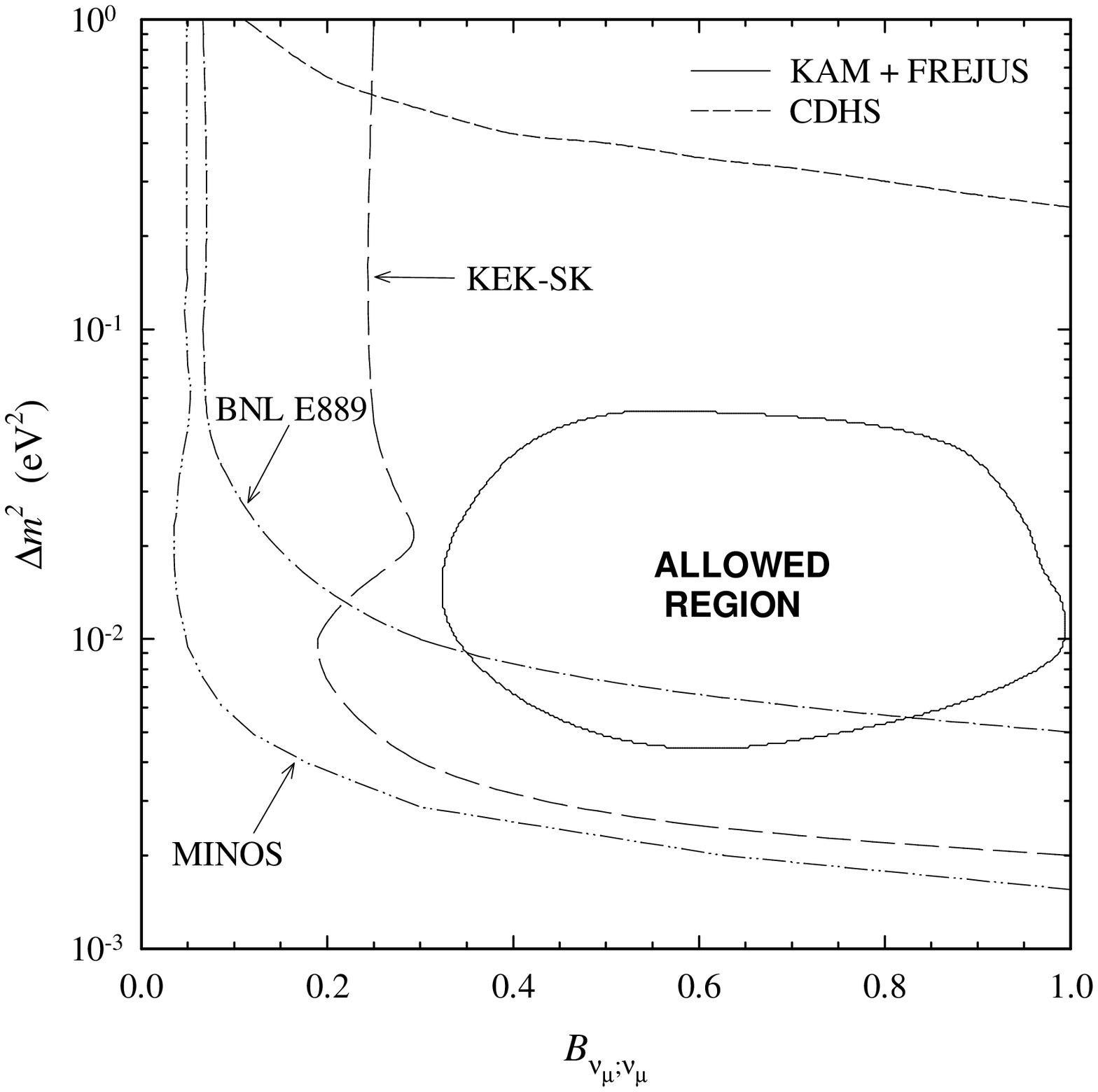,width=\textwidth}}
\end{center}
\vspace{1cm}
\begin{center}
{\Large Figure \ref{FIG6}}
\end{center}
\end{figure}

\begin{figure}[p]
\begin{center}
\mbox{\epsfig{file=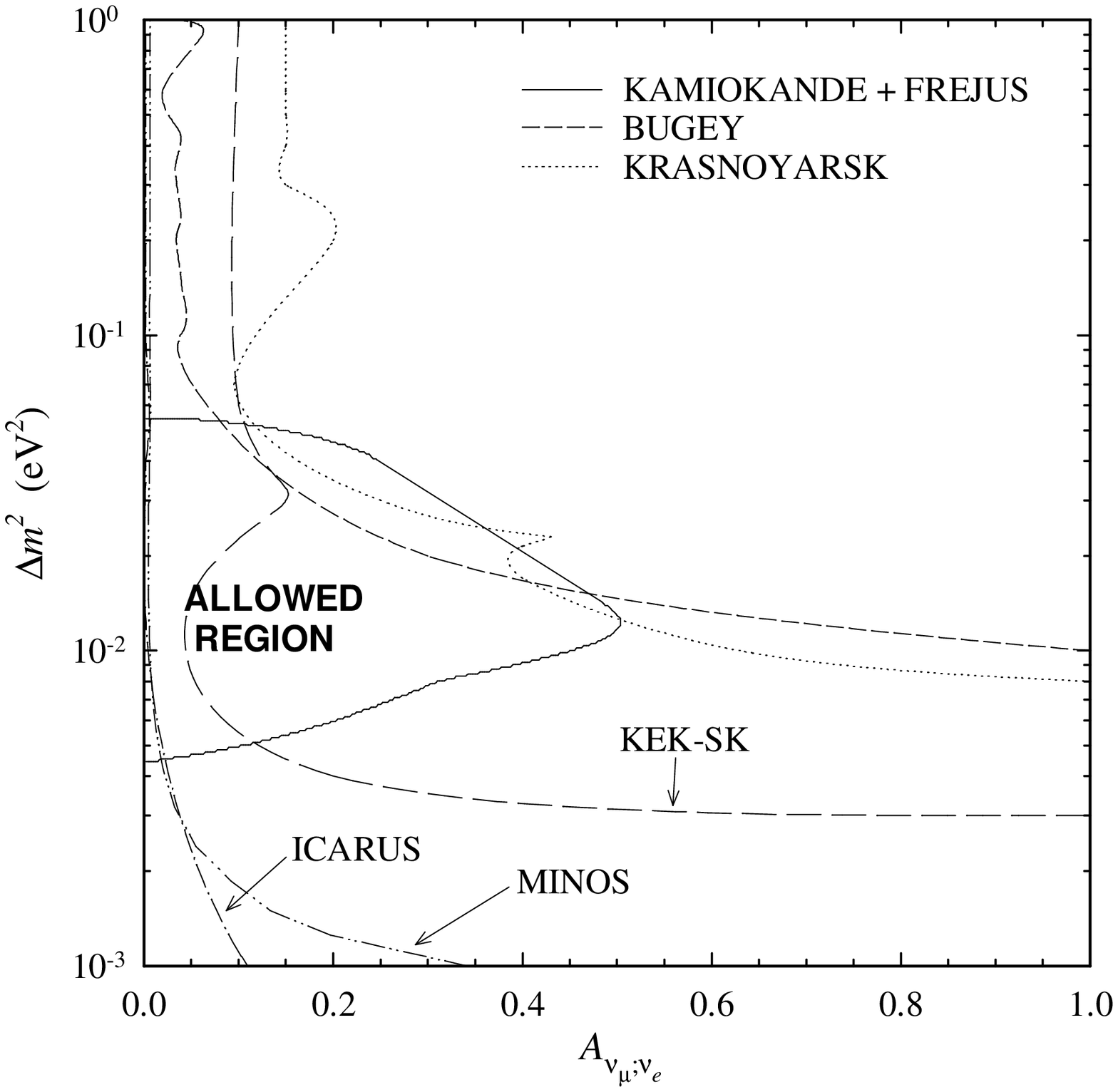,width=\textwidth}}
\end{center}  \vspace{1cm}
\begin{center}
{\Large Figure \ref{FIG7}}
\end{center}

\end{figure}

\begin{figure}[p]
\begin{center}
\mbox{\epsfig{file=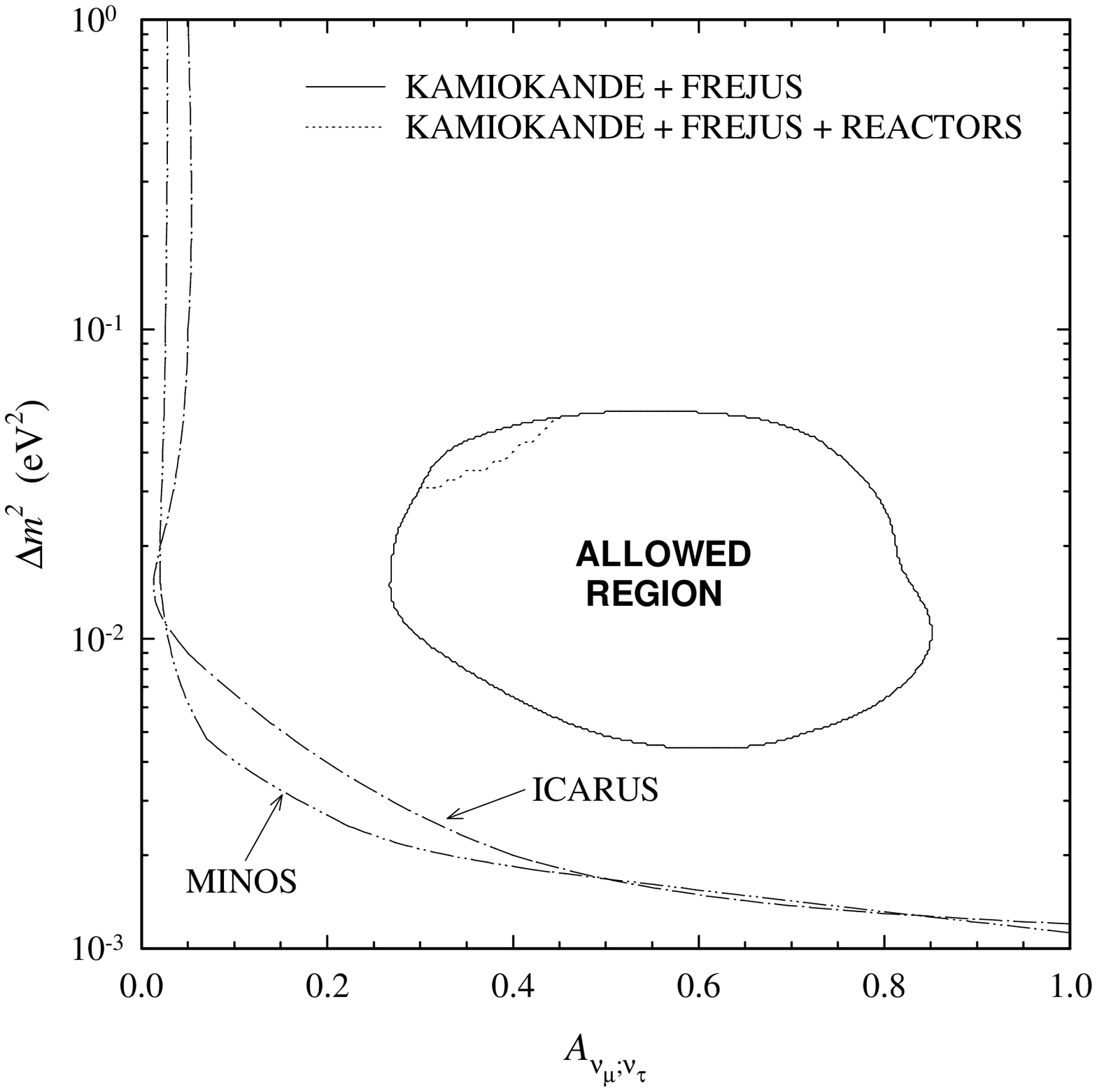,width=\textwidth}}
\end{center}
\vspace{1cm}
\begin{center}
{\Large Figure \ref{FIG8}}
\end{center}
\end{figure}

\end{document}